# A theoretical foundation for building
## *Knowledge-work Support Systems*

Arijit Laha
SETLabs, Hyderabad

June 2008

## *Abstract*

In this paper we propose a novel approach aimed at building a new class of information system platforms which we call the "Knowledge-work Support Systems" or KwSS. KwSS can play a significant role in enhancing the IS support for knowledge management processes, including those customarily identified as less amenable to IS support. In our approach we try to enhance basic functionalities provided by the computer-based information systems, namely, that of improving the efficiency of the knowledge workers in accessing, processing and creating useful information. The improvement, along with proper focus on cultural, social and other aspects of the knowledge management processes, can enhance the workers' efficiency significantly in performing high quality knowledge works. In order to build the proposed approach, we develop several new concepts. The approach analyzes the information availability and usage from the knowledge workers and their works' perspectives and consequently brings forth more transparency in various aspects of information life-cycle with respect to knowledge management. KsSSes are technology platforms, which can be implemented independently as well as in conjunction with other knowledge management and data management technology platforms, to provide significant boost in the knowledge capabilities of organizations.

## Introduction

In recent time *effective knowledge management (KM)* has emerged as one of the most pressing concerns in large organizations. Uses of computerized information systems (IS) for supporting various knowledge management processes have attracted huge attention in both academia and industry. However, till date, majority of the IT-based tools and technologies address the needs of specific activities, typically in stand-alone manner, which form small parts of the larger knowledge-intensive tasks. Very few (if any) IS attempts to integrate the functionalities of these myriad of tools and technologies, often known as the Knowledge Management Systems (KMS), by providing comprehensive organization and ubiquitous access and processing of information resources.

In this paper we argue that to build such systems, we need to concentrate on two basic points. The first of them follows from the fact that *information systems* deal with *information*. Thus, while designing an IS platform for supporting KM we should concentrate on how can we improve the basic and general facilities of access, capture and processing of information. How the information is used is determined by the higher level aspects of the KM processes which are usually decided by the human workers. The second point stems from the fact the information is required by *knowledge workers* to perform their works efficiently. Thus, the design of the information systems must take into account *the information requirement, both by pattern and content, of the workers as they perform knowledge works*. In this paper we build an approach for building IS platforms based on the above which can enhance the capabilities of the knowledge workers by satisfying their information requirement much more efficiently.

Knowledge, according to oxford dictionary is "the sum of what is known". Attempts to define and understand knowledge, at least the documented ones, can be traced back to the classical Greek philosophers Socrates, Plato and Aristotle. Plato's *Dialogue* contains earliest records of the discourses on the subject. For thousands of years thinkers grappled with the issue. Even there exists a rich branch of philosophy, *Epistemology*, dealing exclusively with the nature and scope of knowledge. However, till date no *definitive understanding* of "knowledge" has been achieved.

Despite the lack of the absolute understanding, the importance of knowledge is very much well recognized. Actually, every human activity requires exercise of cognitive capabilities, which in turn, requires information about the environment through sensory inputs and processing of them using knowledge possessed by the individual actor. On the other hand, the very acts of performing activities allow the actor to gain new knowledge and/or update his/her knowledge. Actually, every conscious (some may even want to include subconscious also) moment of a human life adds to and/or modifies the knowledge possessed by a human being.

As long as activities of an individual are concerned, it is clear that the individual is the possessor and user of his/her own knowledge. However, when we turn our attention to "collective activities", where a number of individuals work toward achieving a common goal, the understanding of knowledge-related issues gets complicated. Further, when the activities, individual or collective, are performed while working as part of a larger entity, *organization*, the situation gets further complicated. Actually, discussing knowledge-related issues in organizational setup is akin to stepping into a veritable minefield of debates and conflicting opinions expressed by academic researchers as well as industry practitioners.

Nevertheless, importance of knowledge in organization, especially in postindustrial era (Earl 2001) is well-established. Knowledge is the critical resource, rather than land, machine or capital (Drucker 1993) for success of an organization. However, there is a lack of consensus regarding the nature of knowledge from organizational point of view. Further, large organizations are biggest users of computer-based information systems (IS). So efficient use of IS in context of knowledge has become an important organizational issue (Alavi and Leidner 2001). This led to development of a new discipline of study and practice, Knowledge Management (KM). KM is an area of multidisciplinary study, straddling major disciplines of information systems, management, organizational learning, and strategy (Bray 2007), economics, philosophy and epistemology, computer science and sociology (Earl 2001). KM refers to implementation of various processes, practices and cultures, involving both *social means* as well as use of *information systems*. The KM initiatives can vary widely on several respects across organizations (Earl 2001). Further, the term "knowledge management" is often used somewhat indiscriminately as pointed out in the provocatively titled article "The nonsense of 'knowledge management'" (Wilson 2002).

In the current paper we propose a novel approach for building an information technology-bases infrastructure for improving the quality of IS support available to knowledge workers for performing their works. The approach brings to the center of focus the requirements of individual workers and groups of workers within organization while performing various types of knowledge works. The proposed approach can be used for designing and building a new class of *Information Systems* which we call the "Knowledge-work Support Systems" or KwSS. Here we stress that contribution of our approach is basically confined to the IS usage aspect in KM processes, the other aspects such as cultural and social ones, are not, at least explicitly, influenced by it.

In our approach we do not attempt to *automate* the knowledge works, rather concentrate the effort to capture and deliver useful information *as and when required* to the knowledge workers. We follow the philosophy of Licklider (1960) regarding productive collaboration of human workers and information systems where the best of the capabilities of both can be brought together for accomplishing complex tasks. However, the user can always use the information retrieved from the system to build, maintain and run automated systems such as Decision-support Systems, Expert Systems etc. to get some specific parts of the work done quickly. The KwSS, independently as well as along with existing technology platforms like *Content Management* and *Collaborative Work Tools* customarily used for *knowledge management* in large organizations, can deliver great capabilities to the workers as well as organizations. KwSS can also be used for easily building and maintaining automated systems targeted at specific activities.

To build the proposed approach, we need to examine and understand several fairly complicated issues. They include the relationships among organization, people and knowledge works, the relationships between knowledge and information and their management, the issues of articulation of knowledge and a few more. In this paper we examine these issues to understand various aspects of KwSS. Then we outline the issues addressed by our approach and core features those can be delivered by the systems built based on this approach. Next we propose a model for analyzing the requirements and designing the systems. This is followed by a case study. Lastly we examine the capability of KwSS from the perspective of accommodating various *knowledge management processes* as categorized by two eminent researchers in the field of organizational knowledge management.

## What does an *Organization* require?

Simply put, *an organization needs to achieve a set of goals,* e.g., revenue target, certain market share etc... The requirement of the organization is *to be capable* of achieving these goals. This, in turn, requires high-quality and efficient performance of a large number of tasks. Some of these tasks, usually transactional/operational ones, can be automated to some degree, but the others, especially those involve solving various *semi-structured* and *unstructured problems* (Keen and Scott Morton 1978) require involvement of highly qualified human workers. Such workers are known as "knowledge workers" – a term coined and popularized by Drucker (1959). According to Drucker, a knowledge worker is one *who works primarily with information or one who develops and uses knowledge in the workplace*. Thus, we can say that organization needs to deploy *individual knowledge workers* to get the *knowledge works* performed. Though, as discussed earlier, knowledge is used in every human activity, Mack et al. (2001) defines the term "knowledge work" in this context as *solving problems and accomplishing goals by gathering, organizing, analyzing, creating, and synthesizing information and expertise*. Thus, the capability of getting the knowledge works performed is of *primary importance* to the organization. The scenario is depicted in Figure 1.

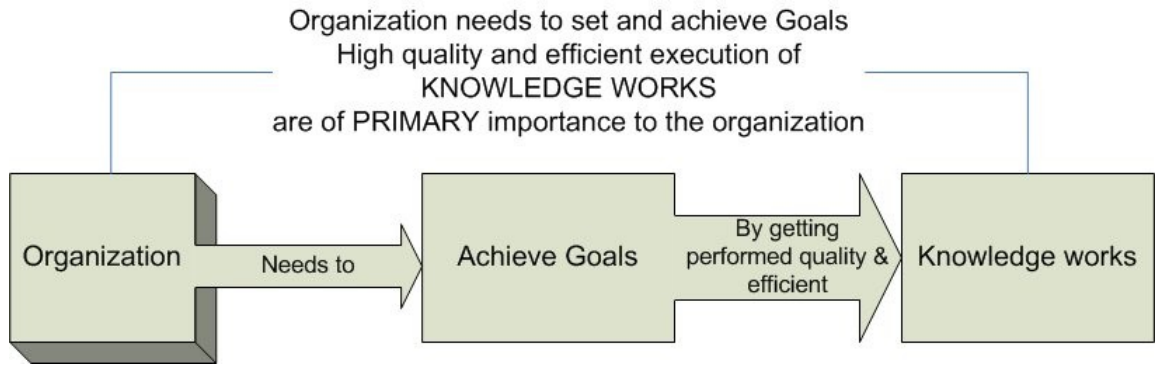

Figure 1: Getting knowledge works performed is of primary importance to the organization

The lifespan of a typical organization is longer than its association with an individual or a particular set of workers. Thus, the need to get the knowledge works performed transcends the availability of a particular set of workers adept at performing the works. Hence, the organization needs to *retain the capability* of getting the works performed irrespective of availability of a particular set of workers. Further, irrespective of whether the available workers are experienced or not, to get a knowledge work performed, the organization must engage knowledge worker(s). However, to achieve quality and efficiency, the knowledge workers need to be supported properly. Figure 2 depicts the interrelationships of various entities and capabilities associated with the performance of knowledge works.

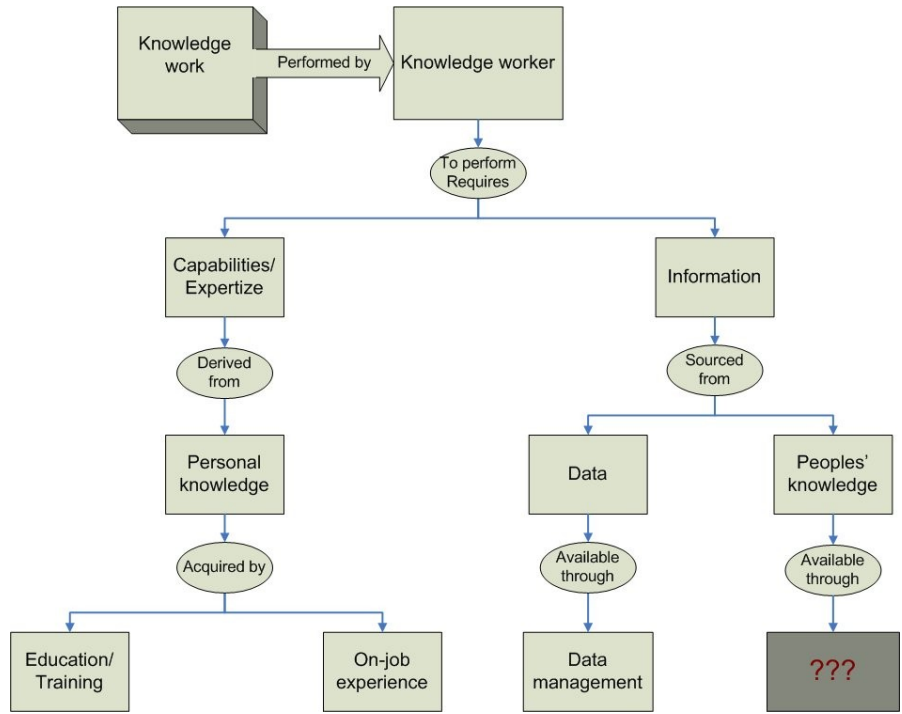

Figure 2

It is clear from Figure 2, that the worker brings to the job his/her personal capabilities and expertise, which are parts of the personal knowledge. Finding and engaging the knowledge workers suitable for performing *target knowledge works* itself is a knowledge work. For the sake of simplicity of discussion, we assume that the management can perform it efficiently. Once the worker is engaged, he/she needs *to access, understand, interpret, analyze* and *synthesize* a *large volume of information* which usually comes from a large number of sources. This results in acquisition of new knowledge by the worker, which, along with his/her existing knowledge, form the body of knowledge required for performing the task assigned to him/her. Thus ensuring the supply of high quality information and providing facility to process efficiently them is organizations' prime responsibility.

The information required by workers can be categorized according to their sources. We can readily identify two sources, (1) data and (2) peoples' or workers' knowledge. Data is consisted of facts and measurements regarding interaction of the organization with external entities (customers, other organizations etc.) as well as internal entities (employees, production facilities, supply chain etc.). Analysis of data produces information regarding the environment. There is already much progress in the area of management and analysis of data, often referred to as Data integration /Data Warehousing and/or Business Intelligence, in organizations.

The other source of information, peoples' knowledge, is consisted of *interpretation* and *contextualization* of the information from *data*, *experiences* and *insights* gained through education, trainings, works performed earlier for the organization and possibly many other activities. It is beyond doubt that availability of such information "as and when required" during execution of knowledge works can make huge difference in terms of both quality and efficiency. The field of study and practice of Knowledge Management (KM) promises to address this issue. In the next section we shall examine briefly some basic KM issues.

In the current paper our main focus is on the issues related to "information developed/derived/sourced from peoples' knowledge" and unless stated otherwise, we shall henceforth use the term "information" to mean the same.

## *Knowledge management*

Knowledge management, as well as knowledge is extremely difficult to define (Earl 2001). Numerous definitions of KM can be found in literature. Prusak (1997) proposed one of the comprehensive definitions of KM as collection of *all the tools, technologies, practices, and incentives deployed by an organization to "know what it knows" and to make this knowledge available to people who need to know it when they need to know it*. Let us examine the definition closely. Firstly, here KM refers to a set of *means* for the organization, for being able to do two things; (1) to "know what it knows" and (2) making the same available to some people at certain time. "Knowing" relates directly to knowledge. An individual is possessor of his/her own knowledge and has a pretty good idea of what he/she knows. Since an organization has to take recourse to a set of fairly elaborate means to "know" what "it knows", what an organization knows or capable of knowing is very much different from the way an individual "knows". Here we argue that an organization as an entity, without considering its people, can not "know" anything (consider the hypothetical situation, in an organization all the people leave at the same time and new people take over. Then what is the use of whatever organization "knows"?). The organization can, at most, be "custodian of information", which enable "people" to form bodies of knowledge, relevant for their works, through accessing, understanding and interpreting relevant parts of the available information. Thus, we can interpret the first part as referring to *the means for an organization to capture "knowledge"* where the term "knowledge" should be understood as *information, whose source is peoples' knowledge*. The second part of the definition deals with utilization of captured information. For utilization, the collected information need to be made available to people:

- *Who need to know it* – the knowledge workers
- *When they need to know it* – while performing knowledge work.

We can add another point, which is implicit in the definition

- *To what purpose* – to get knowledge works performed efficiently, producing high-quality outputs.

Another important and somewhat complementary approach of looking at KM is to consider it as a collection of processes, often called "knowledge processes" and their categorization. Here usage of tools, technologies etc. are considered part of the processes or enabler of the processes, thus making the high level specifications of the processes independent of particular tools or technologies. Such models of KM are called "KM process models". Several KM process models have been proposed by various researchers (Hung et al. 2007). For example, Alavi and Leidner (2001) describe KM as consisted of four categories of processes, Knowledge creation, Knowledge sharing (to include storage and retrieval), Knowledge transfer and Knowledge application. In this model each one of "all the tools, technologies, practices, and incentives" deployed by the organization is part of one of the above categories.

Knowledge processes can also be categorized by whether they involve *knowledge creation* or *knowledge reuse* (Markus 2001). Nonaka (1994) investigated the issue of knowledge creation in organization and introduced the concept of the *knowledge creation spiral* consisted of Socialization, Externalization, Combination and Internalization, known together as SECI (Nonaka and Kono 1998). On the other hand, Markus (2001) developed a theory of "knowledge re-use". Later in this paper we shall examine capabilities offered by the KwSS in light of the SECI model as well as the theory of knowledge re-use.

## Knowledge management and Information Technology

Requirement of "knowledge" is one of the most primary requirements of human civilization. The KM processes or simply knowledge processes, though not always identified as such, have been practiced by human race since the dawn of civilization. At the basic level, *knowledge creation* happens in peoples' mind by understanding, synthesizing and absorbing myriads of information. Knowledge creation, though most important of knowledge process categories, largely utilizes *human cognitive capabilities* and known to be less amenable to technology support, including information technology (IT) (Markus 2001). Most basic means of sharing knowledge is articulation of knowledge by the *knower*. Even in absence of any technology, this can be achieved through verbal means and/or gestures. Use of technology to create the record of articulated knowledge, captures the articulated knowledge as *information*. This, in turn, allows the act of *sharing* to overcome spatial and temporal limitations. Various means such as sending verbal messages by couriers as well as technologies ranging from *clay tablets* of prehistoric ages to *instant messages* supported by modern IT have found their uses over time. Application of knowledge requires human beings to *utilize* their knowledge to *perform* various works.

Thus, the knowledge processes are, in strict sense, are not bound to application of any particular technology. Nevertheless, improvement of technology has played crucial role in terms of improving *ease*, *quality* and *scale* of the knowledge processes. In today's world IT is recognized as a means to provide unprecedented scope of improvements in the above aspects of various knowledge processes. However, the potential is yet to be fully exploited. Markus (2001) observed that *one of the key themes in knowledge management today is the role of information technology (IT) in the transfer of knowledge between those who have it and those who don't*. Information Technology (IT) based systems or simply Information Systems (IS) used in context of KM are popularly known as Knowledge Management Systems (KMS) (Alavi and Leidner 2001). Modern KMSes include numerous applications based on innovative combinations of storage technologies, communication technologies and presentation technologies. Now let us examine them from the viewpoint of the *information requirement of the knowledge workers for performing knowledge works*.

As discussed earlier, to perform a knowledge work, the worker typically accesses a large volume of information from diverse sources. We can divide the sources of information as (1) the people and (2) the archives. The people as source of information include fellow workers sharing the goal, experts and consultants working as advisors etc. They can be thought as the parts of a collaborative effort to accomplish the knowledge work. They *exchange* information among the group. The exchanges of information among the collaborators have several distinguishing characteristics.

1) The collaborator(s) are aware of the source and intended recipients of pieces of information.
2) Significant parts of the information are developed through application of knowledge on-demand by the collaborators, in other words, they are often recent and spontaneous.
3) The exchanges of information are *interactive* in nature, which gives scopes for elicitation of new information from collaborators using the available information as cues.

On the other hand, the archives are the repository of recorded information (which includes documents, manuals, books, records etc.). Here information is *accessed* by the knowledge worker. In contrast with the information from people, they have the following characteristics:

1) The recipients are not known at the time of developing the information.
2) The information in archives usually lack in recency and spontaneity, but they are often of *high quality* due to various corrective and verification steps applied to them earlier.
3) The archives being passive entities, they can not produce any information other than what it contains to begin with.

In case of information exchange among collaborators, use of information technology supports the collaboration to break the boundary of *spatial* and *temporal localization* of the participants. IT based applications for supporting collaboration includes e-mail, chat, conference tools, web-based groupware/collaboration tools and many others. IT is used extensively for archival and retrieval of information. Systems used for this purpose include DBMS, data warehouse (DW), enterprise resource planning (ERP), content management systems (CMS) etc.

## State-of-the-art of IT support for knowledge-works

As noted earlier, in recent time KM has become one of the high priority initiatives in many large organizations. This has created a huge demand for information systems for supporting the KM. There are numerous offerings in the market for systems addressing various activities related to KM. Nevertheless, when we consider enterprise-scale, integrated IT solutions for supporting organizational KM, only a handful of large vendors' offerings stand out. Each one of them typically has started out either as solution for supporting collaborative works or content management. Within last couple of years, they have introduced in their original products features of other class in an attempt to provide complete solutions. However, their core strength still remains in their original sphere of competence. This has led to a very recent market trend where discerning customers in need of cutting-edge functionalities opting for suitably integrated combinations of both classes of products, often sourced from different vendors.

Now we try to understand the extent to which the latest IT offerings help knowledge workers in their works. Note that, IT alone can never script the success of the KM strategies and initiatives. IT, dealing with information, plays the supportive role of an *enabler* only. Other non-IT issues, such as social and cultural, in the organization can have much larger impact in success or failure of a KM initiative. At the basic level, IT, as its name suggests enable the workers to *access* and *exchange "Information"*. Thus, let us examine what support for "accessing and exchanging information" does some of the *top of the line* KMS provide to the knowledge workers. The tools/systems for collaborative works facilitate, as expressed by one of the leading vendors while describes the core strengths of their latest offering as making it *easier than ever to share documents, track tasks, use e-mail efficiently and effectively, and share ideas and information* (Microsoft 2008). The platform also provides *a single workspace for teams to coordinate schedules, organize documents, and participate in discussions—within the organization and over the extranet*.

Similarly, one of the leading content management products is described by the vendor as *a platform that provides a unified environment for storing, accessing, organizing, controlling, retrieving, and delivering any type of unstructured information within an enterprise* (EMC 2008). Here types of unstructured information include text documents, images, audio clips, video clips, email messages, Web pages, XML-tagged documents etc.

Now, let us consider a worker in an organization tasked with performing a knowledge work. We assume that she has been provided with state-of-the-art content management and collaborative work systems. In such a scenario, her capabilities regarding accessing archived information mainly include:
- Use the search facility for querying the CMS
    - Retrieve links to a set of contents matching (predominantly keyword-based) the query.
    - Retrieve one or more actual contents/documents using the links.
        * *Study large portions, often whole, of them one at a time to find useful pieces information.*
        * *(If found) understand, interpret, synthesize and absorb them.*
    - Retrieve more contents, may be using links embedded in the current contents.
    - Repeat the procedure with newly retrieved contents.

And her capabilities regarding exchanging information can be described as
- Contact a collaborator.
- Discuss various aspects of the task in hand through
    - Put across arguments for consideration with short text messages
        * *Prepare documents and other contents to demonstrate rationale behind the arguments.*

- o Communicate the documents and other contents
- o Receive arguments (and contents) from collaborators
    * *Examine them by scrutinizing supporting contents.*
    * *Accept/reject/modify/counter the arguments.*
    * *Prepare and/or append to contents to demonstrate new/modified reasoning.*

Without doubt, the above usage of IT systems renders a huge leap in terms of ease of access and exchange of information compared to a *paper based work-environment*. This translates into significantly improved productivity of the workers and agility of the organization.

## *Beyond the State-of-the-art*

The word "improvement" always connotes a relative positioning. So, the next natural quest is about "further improvements" with respect to the present. Possible improvements can take myriads of trajectory. However, in this paper we are concerned in following the trajectory of improving the productivity of knowledge workers through *betterment of the facilities for accessing and processing information*. To achieve this, in this section, we examine the scopes of improvement and outline the basic features derivable from the improvements. The rest of the paper is devoted to working out how such IT systems can be built.

Let us look up again to the capabilities of the worker as described in the previous section. It is evident that not all the capabilities are uniformly supported by the IS. It can be observed the most important of the capabilities (the innermost *italicized* ones) require greatest intellectual effort on the workers' part while receive perhaps least IS support. Actually, the exercise of these capabilities leads to *knowledge creation* and widely recognized, as noted earlier, not very amenable to technology support. Thus, while there has been great technological progress in other areas, similar focus eluded these core knowledge-intensive activities. In this paper we approach the issue of technology support for *knowledge work* somewhat differently from the established KM approach and investigate how the IT support can be extended to knowledge-intensive activities which are at the core of knowledge works.

We start with the following premises:
1. Information Systems deal with *information*, not *knowledge*.
2. *Knowledge* is possessed by *individual knowledge workers*.
3. Performance of *knowledge work* requires *creation, update* and *application* of *knowledge* by the knowledge workers.
4. *Articulation of knowledge* and its *recording* creates *information*.
5. *Creation* and/or *update* of knowledge require *learning* by the knowledge workers.
6. *Learning* is accomplished through *accessing, understanding, synthesis, interpretation* and *absorption* of *relevant information* by the knowledge workers.

The above premises are largely truisms. Nevertheless, we need to keep them in mind for our discussions.

A worker, while performing a knowledge work, works under several constraints. Most vital of them is that of *time*, i.e., a task needs to be accomplished within certain *time limit*. The quality of work depends on *availability* and *utilization* of the available resources within the time available to the worker. Information is the most vital resource to the worker. So an IT system aimed at improving the quality of the knowledge work must focus on (1) availability, (2) utilization and (3) usefulness of information presented to the user. The proposed approach of building KwSS strives to achieve this.

To understand the extent of support customary IT systems are capable of providing, let us consider a scenario where a worker is engaged in performing a *knowledge work*. In almost all non-trivial cases, the whole work is consisted of a set of *activities* to be performed in a certain sequence to achieve the final result. Thus, for convenience of discussion we shall use the term "task" to denote the larger unit of knowledge work, while use the term "activity" to denote the smaller, constituent units.

At any given moment, the worker typically concentrates on one activity and his information needs are largely focused on those have usefulness in the current activity and other contextually related information. The worker can search and receive the archived information from CMS. Leaving aside other debatable issues regarding whether the documents/contents indicated in the search result contain all the information the worker may need, to gather the *required information* the worker has to read/study large parts, often whole, of the documents/contents one after another. This puts a lot of demand on cognitive ability of the worker as well as the time available. Again, while collaborating with other workers, despite using advanced groupware tools, if the worker tries to explain some points in detail, he/she needs to create new contents, often from the scratch. At the other end, the collaborator, when he/she receives the content, he/she has limited system support to independently verify the premises of the arguments/hypotheses made within content are based on.

Now, let us consider another scenario.
- The worker makes a search for information relevant to the current activity.
    - The system returns pieces of potentially useful information, which are
        - *Not embedded within larger documents/contents, rather extracted from many of them based on the search query as well as activity context.*
    - The worker can study many of them in relatively short time.
    - He/She selects some of them which are found to be of more use than rest.
        - *Study commonalities and differences among them, compute statistics.*
        - *Browse/navigate to other pieces of information related to the selected pieces, individually or collectively.*
    - Quickly gathers enough information relevant to performing current activity, understand, collate, interpret and absorb them.
        - *Gains relatively easily and quickly enough knowledge required for performing the current activity.*
    - Apply the knowledge and performs the activity.

- The worker works with a collaborator.
    - Thinks of communicating a point/argument/suggestion to the collaborators
        - *Quickly gathers enough information relevant to the point under consideration/deliberation, analyze and interprets them and draws conclusions.*
        - *Easily build an information package consisted of the gathered information, the analysis, interpretation and conclusions along with links to sources from which the information is gathered.*
    - Communicate the package along with the argument to the collaborators.
    - The collaborator(s) can
        - *See the argument, find the supporting information and line-of-reasoning on which it is based.*
        - *Can quickly verify the validity of the supporting information by using, if required, the links to sources of information provided in the package.*
        - *Can look up quickly more information independently and/or using various contextual links from the information already available, add them to the information package, demonstrate rationally why he/she accepts/rejects/modifies the argument.*
    - Share the new/modified information package with the group.

To the best of the author's knowledge, as of date, no IT system exists which can comprehensively support the above scenario. Our study suggests strongly that building such IT systems will require looking into a few fundamental issues from a new perspective. In the current paper we attempt to do the same and based on the findings, build a theory as well as roadmap for building a new class of systems, called the "**Knowledge-work Support System (KwSS)**". These systems can be used independently as well as in conjunction of existing KMSes to deliver the required capabilities.

## Knowledge-work Support Systems

Success of the KM strategies adopted in an organization is reflected in the *efficiency* of its workers to perform knowledge works and the *quality* of the products of the works. Thus, if we can consistently and successfully strive to improve and maintain the efficiency of execution and quality of various knowledge works, the organization is likely to achieve its goals. To build successful systems for supporting the knowledge workers in their work, we need to build a deeper understanding of some issues.

### *Knowledge and Information*

Davenport and Prusak (1998) contend that knowledge, though related to both information and data, itself is neither of them. They also observe that the confusion over understanding of natures of these entities has resulted in huge expenditure by organizations in technology initiatives without achieving desired results. They further aver that "minds at work" are the sources of knowledge. Nonaka (1994) opines that knowledge is "justified true belief" held by individuals. Satyadas et al. (2001) recognizes knowledge as a fluid mix of several components which provides a framework for evaluating and incorporating new experiences and information. The components include framed experience, values, contextual and actionable information, and expert insight. These views suggest that the knowledge is primarily an *attribute of individuals acquired through learning and experience which involves their deepest intellectual and cognitive capabilities*. Thus, sharing of knowledge, even partially, to another individual or a group is universally recognized as an extremely difficult problem.

On the other hand, information, though a term commonly used to denote a very large number of concepts, we adopt the view advanced by Davenport and Prusak (1998) that data is *a set of discrete, objective facts about event. Data describes only a part of what happened; it provides no judgment or interpretation and no sustainable basis of action. Data says nothing about its own importance or relevance*. On similar line Nonaka (1994) states that information is collection of *messages from which knowledge can be derived* by a person using his/her cognitive capabilities. Thus, we can identify information as a medium, amenable to recording, preservation and distribution, which help in knowledge creation and re-creation in human mind.

There is a hierarchical relationship among knowledge, information and data. Usually *data,* representing the facts, is given primary importance from IT perspective, which is structured to form *information* and interpretation and contextualization of information is considered *knowledge* (Tuomi 1999). However, the idea of an inverse hierarchy where "knowledge" assumes the cardinal position makes much deeper sense. The inverse hierarchy (Tuomi 1999) follows naturally from the observation that existence of knowledge is primary condition for formulating information as well as measurement of data to form information. Knowledge influence the identification of the parameters to be measured and how the measured values or "data" to be used to form the information. According to this hierarchy, Alavi and Leidner (2001) contend that knowledge *exists which, when articulated, verbalized, and structured, becomes information which, when assigned a fixed representation and standard interpretation, becomes data. Critical to this argument is the fact that knowledge does not exist outside of an agent (a knower): it is indelibly shaped by one's needs as well as one's initial stock of knowledge.*"

Alavi and Leidner (2001) also make crucial observations that information gets converted into knowledge when it is processed by the minds of individuals and conversely, knowledge turns into information when it is *articulated* and *presented* in form of text, graphics, words, or other symbolic forms. They also aver that in order to the knowledge gained by different individuals from same information, be somewhat similar, there must be significant commonalities in their prior knowledge. We can also say that the knowledge derived from certain piece of information, depends on the mental models of the person receiving the information, where mental models, as defined by Steiger and Steiger (2007) are consisted of *hypothetical knowledge structures that integrate the ideas, assumptions, beliefs, facts and misconceptions that together shape the way an individual views and interacts with reality*.

We can summarize from the above views and other discussions as follows,

- Knowledge is created, maintained and applied by individuals' mind through cognitive exercises.
- Individuals can *articulate* some portion of their knowledge. Articulation itself is a *cognitive exercise*, which can take myriads of forms including verbalizing, gestures, writing, painting even literary compositions and artistic performances.
- *Information* is the product of articulation. The recipients of information exercise their cognitive faculty to understand, interpret and absorb the information individually.
- The acts of cognitive processing of information by a recipient add to and/or modify the knowledge *already possessed by the recipient*.
- *Recording* of information by technological means into some persistent media allows us to archive and distribute the information beyond the immediate limits of *locality* and *time* where original acts of articulation take place.
- Modern digital computers and other related technologies can efficiently *archive* and *distribute* digitized forms of recorded information, commonly known as *digital contents* or simply *contents*.
- To support workers in knowledge works the computer must be able to *automatically manipulate* the contents. However, capability of computers in this regard varies considerably across the types of contents. For example, computer systems can store and distribute digitized images or movie clips, but, as of date, can do precious little in terms of automated identification of what is within them.
- *Documents* are a type of contents; in which information are recorded using *linguistic* and *symbolic* constructs are amenable to manipulation by modern computer systems with *reasonable degree of success*.

The above observations will dictate some of the vital characteristics of the KwSS. However, before going into them we need to understand the nature of knowledge with respect to its articulation property.

## How much of knowledge can be articulated?

Michael Polanyi, eminent thinker and philosopher of science, famously observed "…we can know more than we can tell" (Polanyi 1967). He introduced the idea of *tacit* knowledge and *explicit* knowledge largely based on the findings of *Gestalt psychology*. The idea has been of immense influence in the field of knowledge management. There have been a number of different interpretations of this idea. For our purpose we adopt the following:

> "…[there is] a knowledge continuum between the two extreme states of tacit and explicit knowledge, the first being merely in the brain of people and sometimes even hard to explain or to put in words and the second codified or at least potentially codifiable." (Schuett 2003)
> 
> "Knowledge exists on a spectrum. At one extreme, it is almost completely tacit, that is semiconscious and unconscious knowledge held in peoples' heads and bodies. At the other end of the spectrum, knowledge is almost completely explicit or codified, structured and accessible to people other than the individuals originating it. Most knowledge of course exists between the extremes." (Leonard and Sensiper 1998)

From the above we can easily form the idea of articulacy of knowledge that can be represented as a continuous spectrum, ranged between two extremes "tacit" and "explicit". Let us call it the "Knowledge Articulacy Spectrum" or KAS in short. Figure 3 depicts the idea.

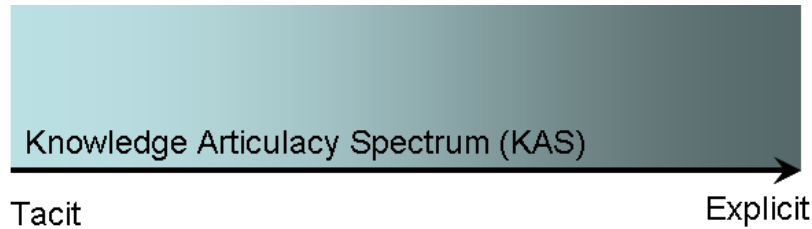

Figure 3

Now, let us remember here that the success of articulation lies in *how useful the 'articulated knowledge' was found by the recipients*. Thus, we can think of a *hypothetical* limit in KAS beyond which the usefulness of the articulated knowledge gets diminished significantly for the recipients. Let us call the limit "Reasonably Useful Articulacy Boundary" or RUAB. Figure 4 depicts the idea of RUAB. Note that, I introduce the concept of RUAB for convenience of discussions only. Determination of such boundary, if at all possible, shall be in the realm of *psychology*.

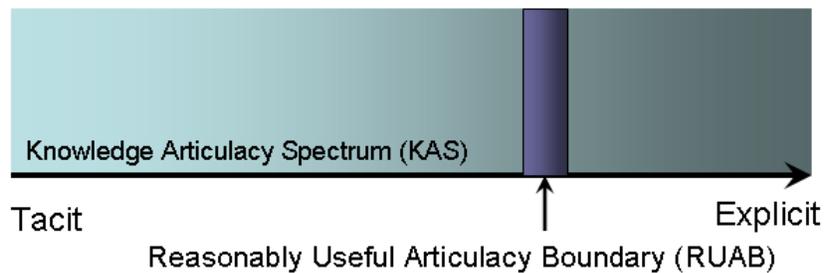

Figure 4

However, it can be easily reasoned that the position of RUAB depends on several factors, including personal as well as organizational. It obviously depends on the intellectual and communication capabilities of the individual as well as the available means of articulation. Further, at organizational level, factors like culture of sharing and appreciation, tools such as brainstorming, discussion, interview, reporting used for knowledge exchange, available IT support etc. may influence the positioning of the RUAB.

From another viewpoint, Snowden (2000) based his ASHEN model on the argument that the knowledge comprised of multiple components (Figure 5(a)). The model supports the idea that it is easier to optimize KM strategies if they are focused on particular components of interest rather than the whole. This view enriches, rather than discards the above if we accept the observation of Tsoukas (1997), which is again largely based on

Polanyi's works, that *all knowledge has necessarily the tacit and explicit parts*. Thus, each of the five components of knowledge in Snowden's ASHEN model has parts those falls along the KAS (Figure 5(b)).

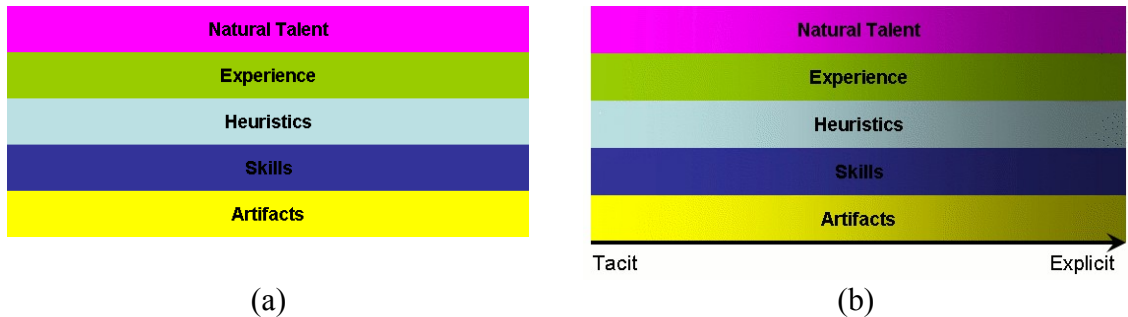

(a)                  (b)

Figure 5

Looking at the components it can be readily recognized that the components may differ considerably in their articulacy potential. For example, the *artifacts* component has very high articulacy (in fact, it may not be considered knowledge at all, it fits better to our description of information), while *natural talents* (say, *of riding a bicycle*) is deeply embedded in individual's brain and enormously difficult to articulate. The situation is depicted in Figure 6. Note that, the relative positions of RUABs for the ASHEN components shown in Figure 6 are purely for explaining the concept we are developing, no concrete meaning should be attached to them. We find the combination of the concepts of components and the KAS together more powerful than any one of them separately.

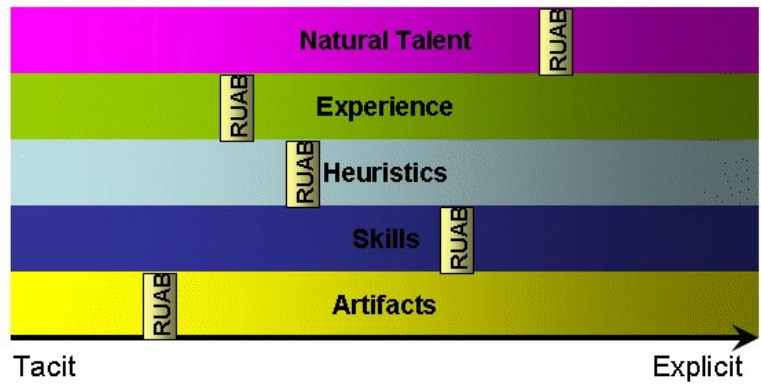

Figure 6

We call this composite model of "knowledge articulacy" the "Components on KAS" (CoKAS) model. We adopt the CoKAS model for following discussion. *Note that, we do not argue that the ASHEN component model 'in particular' is correct or otherwise, rather We contend that the knowledge should be thought as having components with*

*markedly different properties in many respects, including that of articulacy.* Identifying the best possible component model, if exists, is outside the scope of the current paper.

## *KwSS* and the issues to address

As emphasized earlier, we approach the issue of building IT-based support systems for performing knowledge works in a bottom-up fashion. It is most important to observe that the system is used by the worker while performing knowledge work with the aim of best possible execution of the work within least possible time. From this viewpoint we can categorize the tools and techniques used in our approach in two groups. The first group target at ensuring the *adequacy of available information* with respect to the requirement, while the second group address the issue of *information delivery* as and when required.

### *Addressing the information availability*

To understand the information requirement of the worker, we must analyze the same from the perspective of the knowledge work itself. Given a *knowledge work type,* let us call it the "target knowledge work" or TKW, to be performed; it is not difficult to work out the *types of information*, which are required for its quality execution. Once the requisite types of information are identified, the next step is to ascertain *whether the information is available?* If the answer is *no*, we must ensure that the information is made available.

Information originates from the "articulated knowledge", which in turn, produced as intermediate or final products of knowledge works of various types. Thus, analyzing availability of information first leads us to analyze and identify the knowledge work types, which are potential sources of the required information. Let us call them the "source knowledge works" or SKWs. SKWs can be of same type as the TKW as well as of other types, including preparation of essential artifacts such as policies, guidelines, best practices, various reports and analysis etc. To illustrate the point, let us consider as TKW the task of strategic decision-making in business. For such a work, one would like to access information regarding past instances of similar decision-making tasks, as well as information regarding organization policies, market analysis reports etc. Thus here SKWs include similar decision-making tasks and other knowledge works whose products are the policies, market analysis and other informative artifacts.

Now, after we identify the SKW types for the given TKW and ensure that their performances, as required, are practiced by the organization, the next step is to ensure that the workers involved in performing *instances* of the SKWs, articulate enough of the knowledge gained in the processes of performance and the same is recorded as information. The requirement usually spans across several types of knowledge. Here, we need to apply the CoKAS model discussed earlier, to understand the *component structure* of the knowledge types, i.e., the articulacy properties of the useful knowledge types, and devise/optimize the means and methods of articulation of the required knowledge, so that recording of the same can provide adequate information for consumption of the workers involved in performing TKW instances. These steps, when driven from the perspective of

"performing the TKW instances" can lead to availability of enough information satisfying the requirement. We can archive the information at this stage available in form of "recordings of the articulated knowledge". These are, in (if originally not so, been converted to) digitized electronic formats, known as "digital contents" or simply "contents". Types of contents include text documents, images, audio and video clips etc. Content Management Systems can be used for this purpose.

### *Delivering the information to the knowledge worker "as and when needed"*

The knowledge worker needs the information while he/she is engaged in performing a knowledge work *instance*. Let us assume *all the information he/she might need is available among the archived contents*. Now the challenge is to enable the worker to utilize them *effectively*. The first step toward utilization is to retrieve the information as needed from the archive. In current scenario, IT tools allow him/her to search archives for information. Leaving aside the issues of search query formulation and the search algorithms used, what the search usually returns is a set of references to actual contents, for simplifying the discussion let us assume the contents are text documents (same arguments apply to other types of contents also), which potentially contain required information. The next step for the worker is to choose all or a few of the references and retrieve the actual documents for getting the information by reading them one after another.

Under this scenario, the worker retrieves the *whole* documents, in same form as they were originally created. To extract the required information from a document, the worker needs to read *a significant portion, often whole of the document*. However, the portion of the document, containing the relevant information is often *much smaller* compared to the whole document. Thus, to extract enough information from a set of documents, the worker needs to study each document one at a time, sift through many other pieces of information and find out the pieces useful in *current context* from each document, synthesize them to interpret and understand the whole body of information extracted from the set of documents. The whole process is highly knowledge-intensive and given the time constraints, makes a *huge* demand on the cognitive capabilities of the worker. Unfortunately, to the best of our knowledge, no IT-based system exists to help the worker significantly at this stage. In the following we shall investigate the nature of information requirement in this process and define some crucial functionality those can be provided by an IT-based support system in this regard.

### The "context" of information requirement knowledge work

To understand the nature of information requirement during a knowledge work, we must recognize that the information requirement, at any given time, depends on the "context" of the work being performed at that time. An instance of a significant knowledge work is not a *monolithic* affair. To perform a knowledge work instance, we need to decompose it into a sequence of interrelated parts. Let us call a knowledge work instance a *task instance (TI)* and each of its parts an *activity*. Even, a set of TIs can be part of a lager unit of collective knowledge work, which I call an *initiative*. Figure 7 depicts the idea of the se hierarchical units of knowledge work, namely *initiative, task instance* and *activity*. At any given moment, a worker is usually involved in performing an *activity* and the information required by him/her is governed largely by various aspects of the activity. Thus, the context of the *information requirement* at any given time is largely the *activity* being performed at that time by the worker.

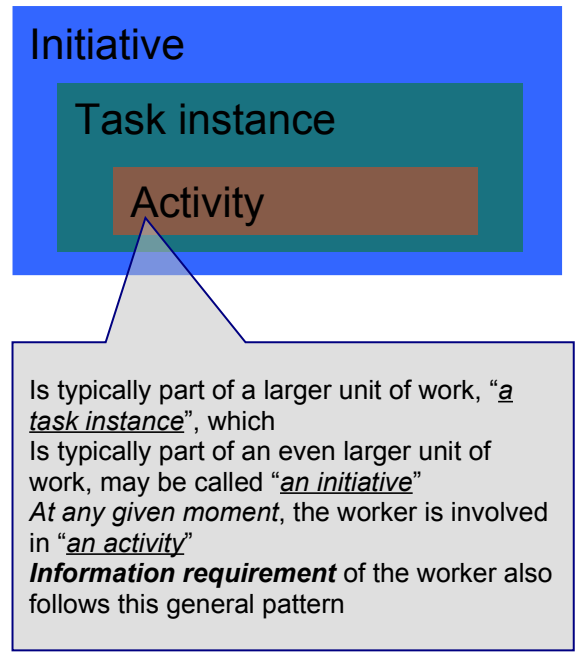

Figure 7

## *An example*

Consider a doctor working in a hospital. The type of knowledge work she mainly performs is to *treat patients*. For her, treating a *particular patient* constitutes a task instance (TI). The TIs can be thought as parts of a larger unit, the *initiative* for providing *medical care*. Naturally, a TI is constituted of a collection of *activities*, starting from collecting the symptoms, signs and history of patient to form an initial impression of possible conditions. This is followed by performing examinations, tests to verify the possibilities, diagnosis of the condition, planning and applying treatments (including medication therapy etc.), overseeing the effect of treatments, adjust treatments according to the response etc. Now consider a scenario: patient P, diagnosed with condition C, undergoing treatment regime T, is showing some unusual signs S. At this stage, the *activity* of the doctor is aimed at finding out the explanation/origin of S and modifying T accordingly.

Now, consider the nature of information requirement of the doctor in context of this particular activity. She will be interested in knowing common causes of S. Then she might like to look-up whether S occurs due to C or T or a combination of both. Are there some observed cases of S occurring when T is applied to patients with C or otherwise?

How were such situations tackled? What were the success rates of those approaches? Were there any complications later on? As we can easily see, the queries are mainly based on the requirement of performing *current activity*. However, even with a content archive and state-of-the-art KM tools, where the doctor can search and access the whole documents related to past cases, to extract all the required information from the contents is a formidable job.

## Contents represent the *Articulation View* of information

As discussed earlier, contents are the digitized records of "articulated knowledge". A worker articulates (part of) his/her knowledge during the course of or after completion of the performance of a knowledge work. The body of knowledge, which the worker attempts to articulate, primarily relates to various aspects of the performance of knowledge work instances. Further, the body of knowledge is formed by accessing, interpreting, understanding, synthesizing various pieces of information and most importantly, by assimilating them with the existing knowledge possessed by the worker. The hallmark of knowledge is its holistic and interconnected nature. When the possessor of knowledge attempts to articulate the knowledge, usually he/she puts together many pieces of related information.

For example, consider the doctor in our earlier example is asked to write a report on the treatment of patient P. Naturally, she will put together all information regarding the patients' symptoms, signs, history, other findings, various reasoning procedures followed in making diagnosis and treatment plan etc. in a document reflecting logical and temporal flow of the activities. Such a report is a typical example of a "document".

Typically, a document contains many pieces of information, related to and/or generated by multiple activities performed as part of the whole task. Let us call this *organization* of information in a document developed through "articulation of knowledge" the **Articulation View of information**, which represent the information regarding a whole task or a significant portion of a task.

## The knowledge worker needs *Exploration View* of information

At a given time a worker is usually engaged in an *activity*, and his/her information requirement is guided by the same, rather than the whole task instances. However, the information typically contained in individual documents, i.e., the articulation views of information, usually include much more information than those relevant in context of the current activity. A lot of *time* and *effort* in part of the worker can be saved if he/she can be provided means to access the information relevant to performing current activity.

Consider the scenario, where the worker, based on the query, is presented with relevant pieces of information, extracted across multiple documents. The worker can quickly go through them, select some of them and access more information, *contextually* related to them. Let us call this desirable organization of information the **Exploration View of information**. However, the information is typically created and available in the *articulation view*.

Thus, one of the major problems in building a KwSS is to compute the *exploration views* of information based on the user queries, while information is available in *articulation view*. In the following section we analyze the problem and formulate a solution of the same.

## Bridging the gap between *Articulation* and *Exploration* views of information

The worker, while engaged in an activity, prefers the "exploration view of information". The exploration view of information is characterized by,

- High *relevance* to the current activity,
- Relevant information, sourced form *multiple documents* and *presented together,*
- Navigational links from the presented information to other *relevant* and *contextual* information, which is embedded within one or more source documents,
- Ability to gather more information through navigation and/or further search, using *links* and *cues* available in the presented information.

However, the information is embedded within documents which presents the "articulation view of information", which is characterized by,

- A *single document* may contain information related to *multiple activities*,
- Naturally, their *usefulness* differ significantly depending on the *context of the current activity*, for which information is sought,
- Information relevant to *a particular activity* in a document is typically contained in a *portion, often small,* of the document,
- Each of a *set of documents* may contain within them portion(s) with information relevant to *a particular activity*.

Evidently, there is significant dissimilarity between two views of the information, as depicted in Figure 8. We need to solve the problem of bridging this gap.

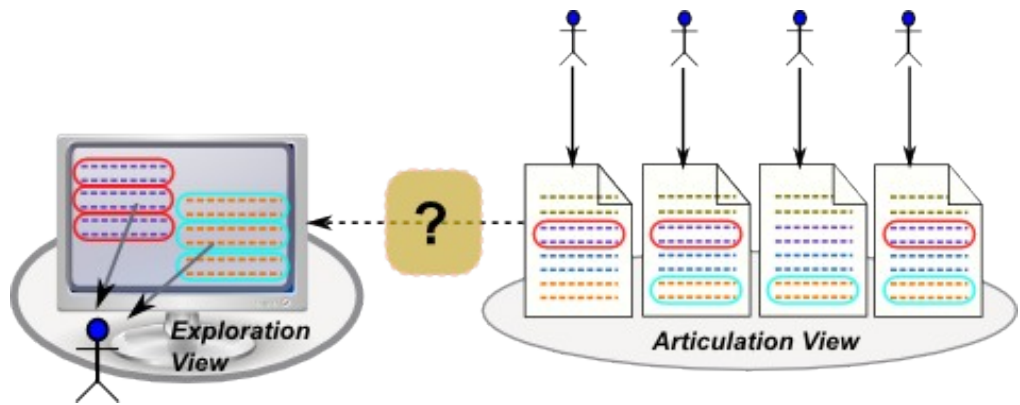

Figure 8

First, let us examine a *lazy approach* to bridge the gap, i.e., we wait till the user asks for some information and only then we try to satisfy the demand. In other words, we attempt to present the worker with the exploration view on-demand by computing it directly from the articulation view. Since the information in the documents is prepared for human consumption, the information is expressed in *natural language* text (except documents with multimedia contents). Building useful exploration views from articulation view in *automated* fashion, requires computer-based techniques of discovering and extracting portions from multiple document according to the requirement specified by the user. Such techniques come under a broad field of research and applications, known as *information retrieval from text* which often employs, as our requirement suggests, *natural language processing (NLP)* techniques to classify/segment the content of the documents into required portions based on various *semantic* and other criteria.

However, using NLP-based IR techniques for our purpose poses two problems. Firstly, NLP techniques are usually *highly computation-intensive*. Thus, unless the number of documents to be analyzed is quite small, the *response time* of such a system could be significantly large. Secondly, as of now, the *accuracy* of NLP techniques, especially when applied to real-life situations are not very satisfactory. Thus, from the perspective of the user, the result produced by the system may not be *reliable* enough to base crucial decisions upon it. Therefore, unless we are ready to employ huge computational resource as well as make significant improvement in accuracy of the NLP techniques, this approach is unlikely to satisfy the user by its *responsiveness* as well as *quality of output*. In the following we formulate an alternative and plausible approach to address these issues.

## *Transcription: the proactive approach*

To circumvent the above problem, we propose a process "transcription" which aims to create an *Intermediate view,* which we call the **Bridge View of information,** of the information which is available in articulation views, i.e., original documents. The bridge view of information will accommodate efficient computation of the exploration views of relevant information based on various types of search criteria. The organizational or structural characteristics of the bridge views may differ depending on the "target knowledge work" types as well as the "source knowledge work" types. We shall explore the issue later in this paper in light of a small case study. Here we shall discuss the methodologies of creating and maintaining the bridge views in general.

Transcription addresses both the issues of *responsiveness of the system* and *quality of retrieved information*. Since the users' information requirement will be satisfied from the bridge view of information, whose design takes into account the users' requirement pattern, the responsiveness of the system will be improved significantly. This is analogous to the situation in "data management" where the operational data is stored in data warehouses in specialized data models which can efficiently accommodate ad-hoc analytic queries. Further, as in ETL processes for data warehouses, quality control measures can be accommodated in the transcription process.

Thus, we can consider transcription consisted of two steps: (1) conversion and (2) verification (quality control). Conversion is the step where information in articulation view is re-organized into the specified bridge view. This step can be implemented, depending on the availability of resources, as manual one or with different degrees of automation. In case of a manual implementation, we can think of the scenario where a human agent, the *transcriber* studies the document, identify various portions of the document and creates through a computer interface the corresponding bridge view of the information in the document.

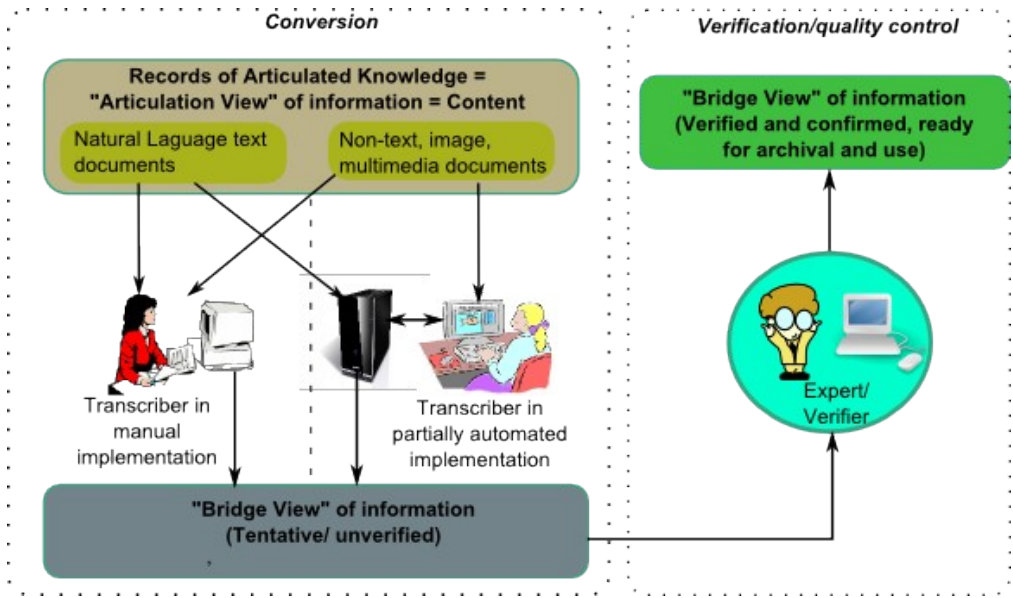

Figure 9

In a partially automated scenario, NLP techniques can be used to analyze small number of related documents and generate a *tentative* bridge view of the information. This tentative version is then inspected and modified by a transcriber, thus making the role of a transcriber more like an *editor*. This would speed-up the process substantially. Further, there system can be designed such the transcribers provide feedback to the system regarding the errors made by the system. The system can use the feedback to adaptively strengthen the NPL technique and improve the accuracy over time. Both the scenarios are depicted in Figure 9.

In the verification stage, the output of conversion step, i.e., the information organized inn the bridge view can be presented to the author(s) of original document(s) or people with similar and/or relevant exercise. They can verify the quality of the converted information, make modification to improve quality and/or assign quality ratings. After that the bridge view of the information can be archived and made available to the users for exploring.

Evidently, the system to be effective, a particular piece of information needs to be available in the "bridge view" before it is needed by the worker. The information has its source in the original documents, which capture the articulation view. Thus transcription of a document needs to be performed some time *after* the document becomes available and *before* the information contained in the document is needed. Smallest time lag in availability can be achieved if the documents are transcribed immediately after they become available. However, depending on the requirement and resource constraints various strategies for scheduling the transcription of the documents can be developed.

# KATE: the model for building *KwSS*

In the previous section we have examine various issues facing the knowledge workers which affect the efficiency and quality of their performance. At the same time we also discussed the solutions for addressing these issues using KwSS. Now we put together the concepts developed till now in form of a model. This model will allow us to study understand various issues related to building KwSS in a systematic manner. It will help us in the requirement analysis and designing actual *Information Systems* implementing the KwSS. The model divides the processes, tools and technology usage into four spaces or categories. They are as follows:

1. **K**nowledge work,
2. **A**rticulation,
3. **T**ranscription and
4. **E**xploration.

Using the first letters, we call it the KATE model. Each of these spaces represents one transformation. Further, the activities within and relationships among them guide us in our effort to design and build KwSS enabled environments. The model is graphically depicted in Figure 10.

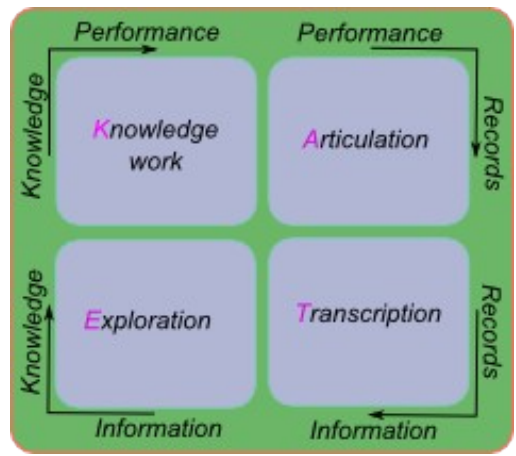

Figure 10

## *Knowledge work*

Knowledge work is the space where the knowledge workers exercise their knowledge to perform knowledge-intensive tasks. In other words, it identifies the processes, tools and technologies which can be used by the workers to *apply* their *knowledge* to *perform* knowledge works. While performing a *knowledge work instance* or a *task instance*, which we call the *target knowledge work* or *TKW*, a worker needs to access, interpret and absorb large amount of information, which in turn is usually created by other workers in past through *performance* of other knowledge work instances, which we call the *source knowledge works* or *SKWs*. To meet the requirement, the worker needs to *explore* information available from various sources. The infrastructure provided by KwSS, aims to allow the worker efficiently access and process useful information. Another common aspect of knowledge work is collaboration among workers. Here workers commonly use sophisticated collaboration tools. However, availability of KwSS will allow the worker to quickly build, exchange and examine information packages relevant to the activities. This will improve the effectiveness of the collaborative works significantly.

The results of the *performance*, from organizational perspective, are solutions to various problems. They may take many forms including a business strategy, formulation of some policies/guidelines, a plan for doing something, an analysis report, a document and many other artifacts. However, along with them, performing the work itself adds to the knowledge of the worker from various sources. This makes him/her potentially more proficient in performing similar works in future. Also, if the worker is able to share the knowledge with others, even partially, they will also be able to improve their proficiency even if they might not have performed the work themselves.

From the perspective of designing KwSS, for a given TKW type, we need to study various aspects of the TKW, including the general pattern of constituent activities and the information requirement for performing them. The next step is to identify the SKWs those can be sources of the required information. We also need to study the activity structure of SKWs and the information generated by performance of those activities.

## *Articulation*

Articulation is the space that addresses development of records of articulated portion of knowledge acquired by the workers in the process of performing the SKW instances. This is the space where we *identify* and *develop* "sharable information. Naturally, the efficacy of the whole system hinges on our success in developing *adequate* volume of useful information. A methodology for working out the requirements for this space can be worked out based on understanding of the CoKAS model of articulacy developed earlier. First we have to identify various components of the knowledge of the workers of SKWs, Note that; here we do not delve into identification of the components in some absolute sense. Rather, we try to identify difference between various parts of the useful knowledge in terms of their articulacy characteristics. For example, the knowledge of a plan formulated can be adequately articulated by written text document, while to articulate how the worker interpreted a set of information as the basis of forming the plan may be more convenient to articulate using a combination of text and diagrams even accompanied with some metaphor or story-telling.

Next we need to study the existing/prevalent practices of articulation and recording of the same for the purpose of sharing information. Such practices can vary widely depending on the requirements felt by the organization and the innovativeness of its management. However, most common of them are the documentation of various kind. They include the final products of the work (strategy, plan, policy etc.) as well as various intermediate products such as reports, memos, messages, consultation notes, minutes of the meetings etc. They may even include logs and transcripts of collaborative works.

Many of these records are often generated routinely and form rich sources of reusable information. However, depending on the information requirement for TKW(s) the KwSS aims to support, these may not be enough. Thus we need to carry out a *gap analysis* and identify the sources of additional information as per requirement. In light of the CoKAS model, we need to identify the types of knowledge those are the potential sources of this information. Then we can work out suitable and effective means of articulation of these knowledge and recording of the same to make the additional information available for sharing. From conceptual standpoint, enabling the articulation of selected portions of knowledge is conceptually equivalent to moving the RUAB(s) leftward for selected *component(s)* of knowledge in the CoKAS model.

Effectively, we have to introduce new processes or modify existing processes and tools for articulation by the workers enabling and encouraging them to articulate the missing portions of required information. However, while doing so, we must attempt to avoid over-burdening the workers to such an extent, that the additional work may hamper the quality of the work itself. For example, in a KwSS for supporting *medical care* in a hospital, the doctors' reasoning processes behind diagnoses or recommendations of treatment can be very valuable information. However, if the existing practice is to write the prescription only, without much details, we might design a process where doctors, verbally dictate the details to a recording device, which will be transcribed later (Actually in several countries such practices exist and known as *medical transcription* which is required by regulation. Nevertheless, the result of traditional medical transcription is to produce *articulation view* of the information, while in our case the transcription refers to production of the appropriate *bridge view* of the information)

## *Transcription*

Transcription, as discussed earlier, consisted of conversion of "articulation view" of information into a suitable "bridge view" of information and verification of the later by experts for quality assurance. Thus, most vital task from the system designer's viewpoint is to design a *suitable* bridge view. The design of the bridge view must strive to meet several criteria. Most important of them is that it should be amenable to computer system based manipulations, so that "exploration views" of information can be computed easily. The nature of the exploration views, in turn, depends on the information access patterns of the workers performing instances of TKW. Further, it also should not be difficult to develop from the information available in "articulation view", which in turn may depend on various aspects of the SKWs and related articulation. Some of these issues are explained later in context of the case study presented.

Once a bridge view is designed, the next step is to develop processes, methodologies and tools for the conversion and verification in efficient manner. The records, as described in relation to "articulation", include final as well as intermediate products of the knowledge works as well as other articulated knowledge. It is possible that the information contained in a set of records may lack in organization and clarity, may have duplication and extraneous elements and many other undesirable elements. To ensure the quality of the output of transcription, the methodologies and guidelines must take this issue into account.

In the previous section, we have discussed general outlines of the transcription processes. Such explicit implementations of transcription are necessary in many cases, especially if we want to make the contents of older documents usable through KwSS. However, in many environments where the workers routinely use IT tools for their works, suitable computer interfaces may be provided to them for articulation, which will directly record the information in bridge view, rendering the transcription step implicit. Even with implementations starting with explicit transcription, the system may gradually switch to the implicit transcription with increased maturity of the users in using the system.

## *Exploration*

Exploration includes tools and techniques of searching, retrieving and studying information useful for performing TKW instances. Studying, understanding and interpreting the retrieved information cause creation and/or update of the knowledge of the user, which in turn, increases his/her ability to perform the knowledge work(s) with improved capability. Here the challenge is to understand the information requirement pattern of the worker and design the tools and techniques for search, navigation and information presentation in such a way that the requirement is fulfilled easily. We have already discussed various aspects of this requirement pattern in earlier part of this section.

From the system design perspective we need to focus on providing interactive search and navigation facilities, preferably including semantic and pattern-based features. These will enable the workers to easily find useful information. Also to enable the workers to quickly consume the information, we need to work out various information presentation strategies. These will depend strongly on the TKW as well as several factors depending on the application domain and people. Figure 11 depicts a graphic representation of typical types of actors, tools and systems used in each spaces of KATE.

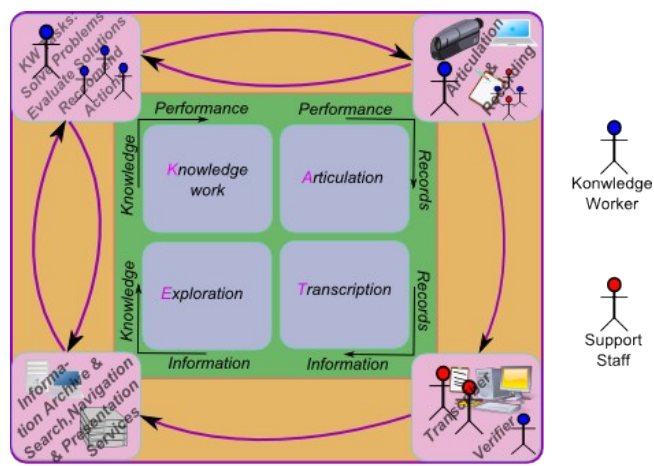

Figure 11

## *Case study: using the KATE model*

Here we explain how the KATE model can be used to study the requirement for building KwSS for a particular type of knowledge work. Here we consider the problem of designing a KwSS for helping knowledge workers engaged in decision-making in business environment, i.e., using our terminology, the TKW for the KwSS is "business decision-making". Laha (2006) proposed the architectural outline of such a system, though not explicitly identified as a KwSS. Here we shall analyze various aspects of the architecture in light of the KATE model.

A business organization is typically in lookout for the status of various internal and external factors impacting the business performance. Measurement/observation of those factors and analyses of them allow the business to detect various *opportunities* as well as *threats*. The organization needs to take actions to take advantage of the opportunities and counter the threats. Formulating an action to take, which is the core of the TKW here, in turn, requires good understanding of the situation, possible solution alternatives and their relative merits, choosing an optimal solution, prepare the plans for implementation etc. All these can be considered as various activities constituting the task of decision-making.

Now, to understand the information requirement for performing the TKW, we identify a set of questions whose answers can be of great help to the worker. Some of those questions are as follows:
- Were there similar situations those the organization faced in past?
- What were the responses then?
- Why was that particular response chosen for that particular situation?
    - What were the alternative solutions?
    - What was the desired outcome/impact of the chosen path of action?
    - What was the actual outcome/impact?
    - Why did they differ?
        - Problem was with the solution chosen or implementation?
        - Were there some other factors those were not accounted for?
    - What might happen if a similar decision is taken now?
        - How do the business conditions differ?
        - How do the market conditions differ?
        - What the relative positions are in terms of scopes, deployable resources etc.?

Figuring out how such information can be sourced, leads us to identification of the SKWs, which will include here past decision-making tasks themselves, the implementation of the plans by the operations managers, later analysis of the impact of the actions.

Above analysis addresses the space "K" of the KATE model. Now we turn our attention to "A", where we have to ensure adequate volume of the required information can be made available. In a business organization, the management usually has a hierarchical structure. Therefore, a decision-making task in such environment, involves routine overview and scrutiny by peers and supervisors. Thus, more often than not, the standard practices include good documentations of the activities. Though they, especially the interim ones, are largely meant for facilitating the performance of current task, they can be readily used for sourcing information useful to other workers.

However, the gap analysis reveals several requirements for improving the articulation of the workers' knowledge. Here we describe one of the most important issues considered and how that can be tackled by introducing an additional step in the decision-making process and articulation of the knowledge developed thereof. Usually a decision-making task results in formulation of the decision, a plan for implementation of the decision and an estimate of the impact on business if the plan is implemented properly. The responsibility of implementing the plan usually falls upon a different group of workers. Further, assessment of the actual impact, deviation from estimate and the causal analysis of the deviation is performed much later and often by a third group of workers. This situation often results in inconsistent, incomplete and delayed feedback on the efficacy of the decision, leaving little opportunity of taking corrective measures or fine-tuning the strategies.

To address the problem, we introduce the concept of *watch-plan*. A watch-plan is basically a list of measurable/observable parameters and a schedule for their measurements. The watch-plan is developed by the workers involved in decision-making. The implementation plan for a decision and its estimated impacts form the basis of identifying the parameters whose values will reflect the performance of the business as affected by the implementation of the particular decision. Further, a time schedule can be worked out for carrying out the measurements of the parameters when the impact on the parameter should become evident. A documented watch-plan can be used by the implementers to ensure timely and adequate data capture for computations of the parameters as per schedule. For the analysts, the watch-plan becomes a valuable guideline for detecting the real impacts and deviations from the estimates. Last, but not least, from the KwSS design perspective, the documented watch-plans become important sources of information for other decision-makers.

Transcription deals with the design of a suitable bridge view, processes and tools for creating and maintaining it. Based on the understanding of the information requirement as well as typical patterns of activities, the bridge view proposed by Laha (2008) takes the form of a network/hypernetwork of information. The network is named "organizational experience". Each of the nodes in the network is called an "event". An event object corresponds to an activity in a decision-making task instance. Its content is the information developed by performing the corresponding activity. Events are linked with other events based on their *contextual* and *relevance* interrelationships.

Considering the high level of IT adoption in modern business organizations, here attempt has been made to merge the "articulation" and "transcription", where the workers themselves can be provided with tools and interfaces for documenting information activity-wise directly in form of the *organizational experience*. However, for making the contents of existing (generated and archived before the use of KwSS is started) documents available, an offline transcription facility in the line described earlier may be needed initially.

To satisfy the "exploration" requirement of the workers, we need to provide suitable tools and techniques of search, navigation and information visualization. In context of the current system, we can think of an integrated, interactive tool to accommodate these three functionalities. For searching, in addition to traditional *keyword*-based techniques, *semantic* and *pattern*-based techniques will deliver much power to the user. Further, various *hyper-linked text*-based as well as *graphical* interfaces for information-visualization and navigation will be of great use. Works on developing such interfaces is currently in progress.

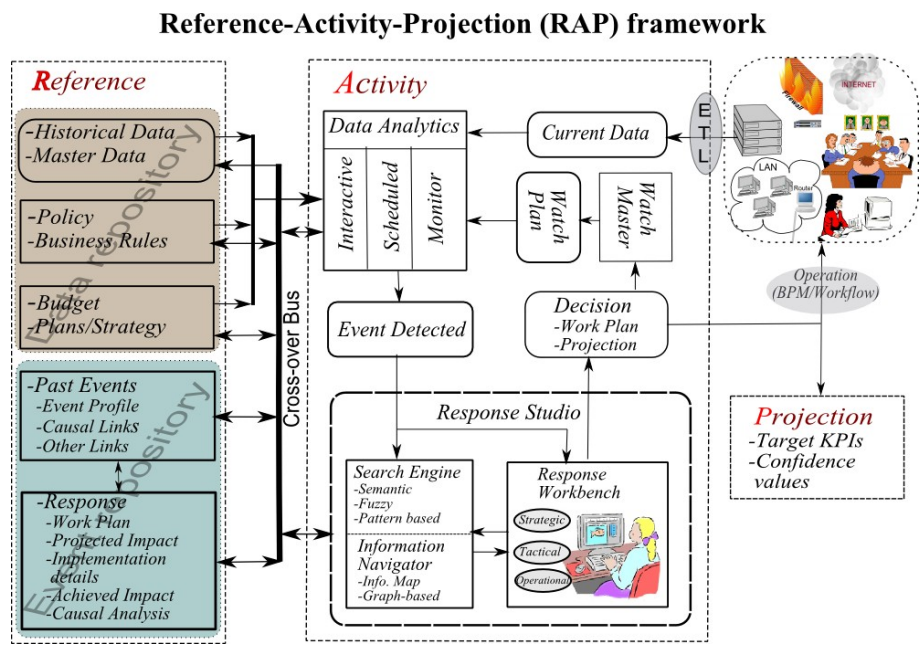

Figure 12

Figure 12 depicts the architectural outline (Laha 2008) of a KwSS environment for business decision-making. The design assumes existence of data warehouse or similar environment for data integration and the KwSS acts as an add-on or value-multiplier. For details readers are referred to (Laha 2008).

## *KwSS* support for Knowledge Management processes

As mentioned earlier, knowledge management in organizations is a vast and all-encompassing enterprise. To effectively implement knowledge management, the organization needs to utilize tools and techniques developed in a large number of fields of study. Information technology is one of them. Though IT plays a major role, it is mainly that of an *enabler*. KwSS provides a technological environment for easy and efficient access of information for performing knowledge works. KwSS can be implemented independently as well as in combination with other IT solutions customarily used in the KM space. We have already seen how it can utilize contents archived in Content Management Systems and turn them into more valuable resource through transcription. We have also discussed the potential of KwSS in improving the efficiency and productivity of collaborative activities using Collaboration Tools. KwSS are *enabler* designed at improving the system support for some basic and general needs of knowledge workers. Here we try to understand from KM perspective how the facilities provided by KwSS can be used to serve various higher level "Knowledge management Processes".

There are various ways of categorizing the KM processes. Here we follow the categorization of knowledge processes into *knowledge creation* and *knowledge reuses* (Markus 2001). These categories can be further divided into subcategories using various criteria. Nonaka, in his famous theory of "Organizational Knowledge Creation" (Nonaka 1994) organized the knowledge creation processes in four categories. In relation to knowledge re-use, Markus (2001) advanced another categorization. In the following we shall examine how the facilities provided by the KwSS can serve the classes of knowledge processes as identified by Nonaka and Markus respectively. Note that, the processes of "knowledge creation" and "knowledge reuse" are not mutually separate. Clearly, to create new knowledge we need to use information which originates from knowledge. On the other hand, reusing knowledge (or using information) causes gain of knowledge by the user, leading to knowledge creation in his/her. These two categorizations should be considered as two alternative expositions of same subject matter.

## *KwSS and Knowledge Creation processes*

From our viewpoint, the knowledge capability of an organization reflects the resources available to the organizations for getting various types of knowledge works performed, which are required for achieving the organizational goals. From this perspective, *knowledge creation in an organization* can be viewed as (1) some workers gaining knowledge required for new types of knowledge works, (2) more workers gaining capabilities of performing certain types of knowledge works which might have been held by a smaller number of workers. Knowledge creation processes practiced in an organization caters to both the needs. Nonaka (1994) distinguished four categories of knowledge creation processes, *Socialization, Externalization, Combination* and *Internalization*, often referred together as the *SECI model of knowledge creation*.

### Socialization

Socialization refers to the processes of sharing *tacit knowledge* between individuals (Nonaka and Kono 1998) and considered to be least amenable to technology support. However, *tacit knowledge* is, by definition, deep-rooted in the minds of the individual possessors and can not be *shared* in its original form. When someone wants to share his/her tacit knowledge with another person, he/she articulates the knowledge as best as possible. The articulated knowledge is *information*, which is consumed by the other and assimilated with his/her existing tacit knowledge, thus gaining tacit knowledge.

Therefore, the *socialization* between two individuals is a *two-way process* as depicted in Figure 13. Both the parties in the interaction produce information through articulation of their knowledge and absorb information produced by other. In case of more than two participants the same model extends to include consumption of same information (result of articulation of one participant) by multiple participants.

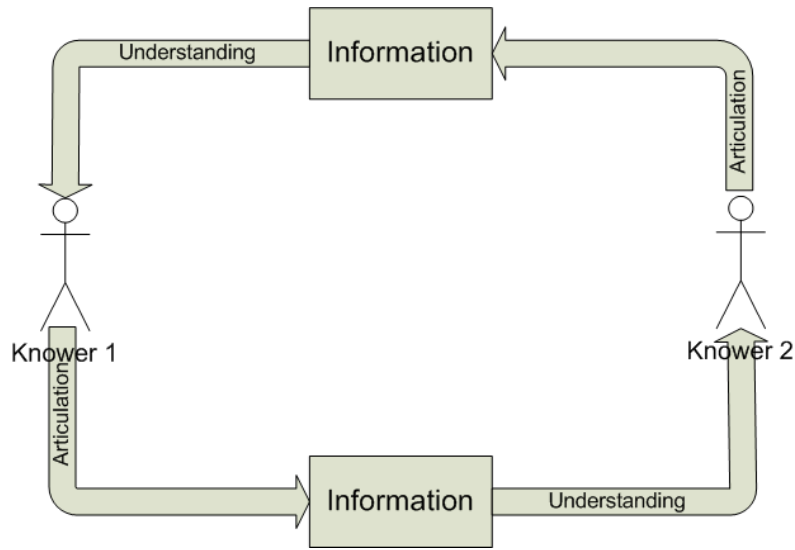

Figure 13

Though in socialization, the individual participants and their knowledge form the most important component, it can be observed that *information* acts as the intermediary of the knowledge exchange process. Thus, the better is the quality of the information produced and consumed in the process, more productive and efficient it is. KwSS can enable the participants to produce high quality information by using the facility of quickly searching and assembling relevant information to value-add/support the information which might have been produced by articulation of knowledge alone. As far as the information consumers are concerns, he/she receives more comprehensive information and system support to help processing as well as verification the information received. This will ease the process as well as extend the capability of absorbing the information and gaining/updating tacit knowledge. The scenario of KwSS enabled socialization process is depicted graphically in Figure 14.

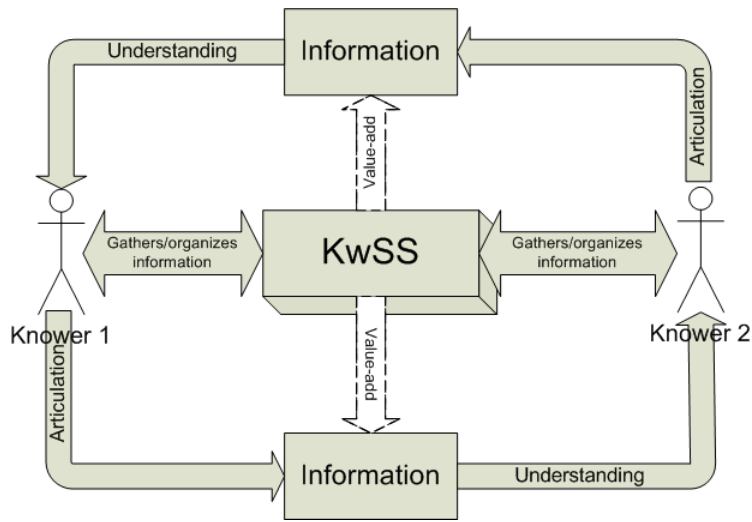

Figure 14

## Externalization

Nonaka (1994) contends that the concept of externalization is far from well-developed. Nevertheless, we can look toward the processes in this category as articulation of knowledge by one or more persons to an audience. Here the difference is that the audience only consumes the information, they do not produce information for consumption of the producers. Therefore, it differs from the socialization by being a *one way* communication. Thus, if the articulated knowledge i.e., the information produced is *recorded* and *archived*, it can be made available to the audience beyond the time and localization of the occurrence of the act of articulation. The benefits of KwSS environment for such processes are similar as socialization. It provides support for efficient production of rich information and support for easy processing, verification and consumption by the audience.

## Combination

Combination represents conversion of set of *explicit knowledge* to into other form, often *more complex* set of *explicit knowledge* (Nonaka and Kono 1998). They involve reconfiguration of existing information by various means including sorting, adding, re-categorization and re-contextualization to create new information. Naturally, these processes can benefit hugely from application of information technology. Accordingly there is good adoption of *information processing* tools to address various requirements of such processes, often automating them substantially.

However, we must recognize that any non-trivial combination process is knowledge-intensive and needs human expertise to be performed. It requires human contribution in terms of selection of source information, understanding and interpreting them, directing the processing and synthesis of the source information to produce the desired information. From KwSS perspective, the role played by the human worker in the combination process itself can be identified as a *type of knowledge work*. Therefore, KwSS can be easily designed to support the worker extensively.

### Internalization

The internalization processes aim at allowing the workers to gain knowledge through learning. Such processes include training, exercises, on-job-learning and many more. The archived information is one of the most important resources utilized in such processes. Naturally, by its design, KwSS core capabilities of efficient retrieval of useful information and presentation of the same in user-friendly manner can be of great use for improving the quality of internalization processes.

## *KwSS and Knowledge Re-use processes*

Markus (2001) identified four types of workers distinguished by their knowledge reuse characteristics. They are (1) shared work producers, (2) shared work practitioners, (3) expertise-seeking novices, and (4) secondary knowledge miners. Shared work producers are collaborators in close group working toward a common goal. They are characterized by possession of similar or cross-functional knowledge. The shared work practitioners usually perform similar works in different settings. They need to use the information produced by others for performing their works. The secondary knowledge miners seek answers to new questions and develop new knowledge by analyzing archived information. It can be readily recognized that support for all these categories can be adequately provided through KwSS.

The third category of knowledge re-users, termed as *expertise-seeking novices* require identifying experts who possess some expertise vital for their works. Actually, the requirement of expertise-seeking is not restricted to the *novices,* rather is a very crucial and frequent problem faced in modern organizations. The problem assumes gigantic proportion in case of large organizations, whose people are based at diverse locations all over the globe. KwSS can help substantially in solving this problem. In KwSS the bridge views of information is archived. As discussed earlier, the bridge views are designed largely based on the activity-structures of the tasks. Thus, in KwSS it is easy to keep record of contributions of workers task-wise as well as activity-wise. Further, contextually linked information can be used to analyze the quality of the contributions of the workers. Thus, we can easily build applications on the top of KwSS to find people with requisite expertise as well as build their performance-based profiles.

# Conclusion

In this paper we have proposed a novel approach for building information systems, which we call the Knowledge-work Support Systems or KwSS, those can aid knowledge-workers in efficiently capturing, organizing, accessing, integrating and processing of *useful* information. Customarily these are identified with "knowledge management". However, in our approach we attempt to enhance basic IS support in a systematic manner for improving the efficiency and quality of higher level knowledge management processes. Our approach makes significant departure in putting the knowledge-works and the workers in the center of focus and the whole system is designed based on the information requirement of the workers while performing instances of some targeted types of knowledge works. It enables us to bring much transparency in terms of identifying various types of information and their usage.

In the course of the development of our approach we have developed several ideas. They include the CoKAS model of knowledge articulacy, the "articulation" and "exploration" views of information and use of "transcription" for building bridge views to overcome the difficulty of computing exploration views of information directly from those available in articulation view. Finally we put these concepts together to build the KATE model, which guides us in requirement analysis and design of system architectures for particular KwSS.

KwSS are technology platforms can be loosely thought of playing a role in *information space* somewhat analogous to the one played by data warehouses in the *data space*. In fact, there are points of similarity between these in the sense that in a DW the source data is stored in the DW data model to enable ad-hoc and advanced data analysis. The "bridge view" of information plays a similar role in a KwSS. In that sense the role of ETL in DW and that of "Transcription" in KwSS are somewhat similar. Nevertheless, the tasks performed by them are actually opposite. ETL sources data from the operational databases where data is in its most granular form, collected through well-defined means including automated ones. Then ETL aggregates those following well-defined mapping rules to load into the DW data model. In case of transcription, the sources are the documents and other contents, which originally contain aggregated information, and more importantly, the aggregations found here are results of *human intellectual efforts* rather than some routine automated process. Transcription is used for analyzing such contents and decomposing them into smaller pieces which will have wider utilization prospect. Naturally transcription is a much more challenging task to implement.

In this paper we concentrated on the information sourced from the workers within the organization. However, that does not exclude the facility of sourcing information from outside, e.g., the Web. We must recognize that the material available in the web is the product of articulation of somebody's knowledge and typically organized as *articulation views of information*. We may not have the power to influence the articulation to get anything more that what is already there, but we can access the web documents directly while exploring, and study them as whole. The other possibility is quite exciting and possible only because of the KwSS. We can collect (may be using some agent/bot technology) a set of potentially useful web documents and redirect then to transcription so that the information from web can be viewed in exploration view along with others.

The other exciting possibility is to interface the KwSS with data integration/management systems. Results of data analysis, when understood and interpreted become very important source of information, especially about the business environment. Now consider the scenario where a worker in a KwSS environment needs to find something from the data, she has an interface through which she can design and launch a data analysis task to be executed through data management systems and get back the results. Then she understands and interprets its significance in lieu with other available information. Next, in corresponding articulation step she includes the details of this data analysis in the information developed. Later, a different worker finds the information useful in general, but the environment might have changed by then. In such situation, she can use the design of the data analysis developed by the previous worker and use it as template to quickly design and fire a new data analysis task to investigate the current situation.

In this paper we have described the basic ideas, from somewhat theoretical standpoint, behind the KwSS. KwSS being a *platform* or *infrastructure* for supporting knowledge processes, there can be various options regarding the use of technologies for implementing them. Currently we are working on designing different technology stacks for implementing KwSS in different environments. We are also in the process of building various frameworks, architectures, specialized tools and technologies related to the same.